\documentclass[longbibliography,12pt]{revtex4}

\usepackage{ragged2e}
\usepackage{graphics}
\usepackage{graphicx}
\usepackage{longtable}
\usepackage{lineno}
\usepackage{float}
\usepackage{amsmath,amsthm,amssymb}
\usepackage{xcolor}
\usepackage{natbib}

\begin{document}


\title{Modeling of radiative emission from shallow color centers in single crystalline diamond}


\author{Maryam Zahedian$^{1,*}$, Jietian Liu$^{1,*}$, Ricardo Vidrio$^{1}$, Shimon Kolkowitz$^{2}$, Jennifer T. Choy$^{1,**}$}

\affiliation{$^*$These authors contributed equally to this work.}
\affiliation{$^1$Department of Engineering Physics, University of Wisconsin - Madison, Madison, WI 53706}
\affiliation{$^2$Department of Physics, University of Wisconsin - Madison, Madison, WI 53706}
\affiliation{$^{**}$Email: jennifer.choy@wisc.edu}

%



\begin{abstract}
\justifying

Optically active defects in diamond are widely used as bright single-photon sources for quantum sensing, computing, and communication. For many applications, it is useful to place the emitter close to the diamond surface, where the radiative properties of the emitter are strongly modified by its dielectric environment. It is well-known that the radiative power from an electric dipole decreases as the emitter approaches an interface with a lower-index dielectric, leading to an increase in the radiative lifetime. For emitters in crystalline solids, modeling of this effect needs to take into account the crystal orientation and direction of the surface cut, which can greatly impact the emission characteristics. In this paper, we provide a framework for analyzing the emission rates of shallow ($<100$ nm) defects, in which optical transitions are derived from electric dipoles in a plane perpendicular to their spin axis. We present our calculations for the depth-dependent radiative lifetime for color centers in (100)-, (110)-, and (111)-cut diamond, which can be extended to other vacancy defects in diamond.

\end{abstract}


\maketitle

\section{Introduction}
\justifying

Color centers in diamond, such as the negatively charged nitrogen-vacancy (NV) centers, are solid-state defects whose spin properties enable sensitive measurements of the local magnetic fields~\cite{hong2013nanoscale,barry2016optical,rondin2014magnetometry}, strain~\cite{doherty2014electronic,ovartchaiyapong2014dynamic,knauer2020situ}, and temperature~\cite{plakhotnik2014all,acosta2010temperature,chen2011temperature} through spin-dependent photoluminescence. For sensing applications, the measurement sensitivity and spatial resolution benefit from placing the spin defect in close proximity (within few to tens of nanometers)
to the physical quantity being measured. Typical sensing measurements thus involve placing color-center-containing nanodiamonds close to the sensing target (via attachment to scanning probes~\cite{schell2011scanning} or spin-casting) or using implanted~\cite{orwa2011engineering} or delta-doped~\cite{ohno2012engineering} shallow color centers in single crystal diamond.

For color centers close to a planar surface, it is important to consider the effect of their dielectric environment on the emission characteristics, in particular the rate of energy emission. There is a host of theoretical~\cite{drexhage1970influence,lukosz1977fluorescence,barnes1998fluorescence} and experimental~\cite{danz2002fluorescence, schell2014scanning} work on the dependence of the emitted power of optical dipoles on their distance to an interface. For emitters in single-crystalline solids, this effect is complicated by the crystal orientation and direction of the surface cut which constrain the relative orientation of the spin and thereby electric dipoles to the surface. Understanding the characteristics of shallow-color-center emission, including how the radiative lifetime is affected by proximity to the surface, is especially crucial for implementing near-field sensing measurements such as F\"{o}rster resonance energy transfer (FRET)~\cite{tisler2013single, nelz2020near, radtke2019nanoscale} and for designing photonic structures to control light-matter interactions with color centers~\cite{faraon2011resonant,hausmann2013coupling, wambold2021adjoint}. 

This paper aims to provide a framework for analyzing the emission rates of near-surface color center in diamond. We primarily focus on the negatively charged NV center (hereby abbreviated as NV) and calculate the radiative lifetimes of NVs close to the air-diamond interface, by integrating their total radiative power. We first summarize and review the relevant literature on the photo-physics of NVs, which inform the treatment of the electric dipoles in our model. We then explain our approach to calculating radiative lifetimes of optical dipoles near surfaces. Calculations of NV lifetimes as a function of depth for diamond with (100)-, (110)-, and (111)-surface cuts are then presented, showing the depth-dependence of the NV radiative lifetime within 100 nm from the interface and the effects of dipole orientation and diamond surface cut. We conclude with a discussion on extending our approach to other color centers and the feasibility of using lifetime measurements to determine emitter depth.

\section{Model of optical dipole transitions in vacancy defects} \label {photo-physics} 
\begin{figure}[b]
    \centering
    \includegraphics[width=\textwidth]{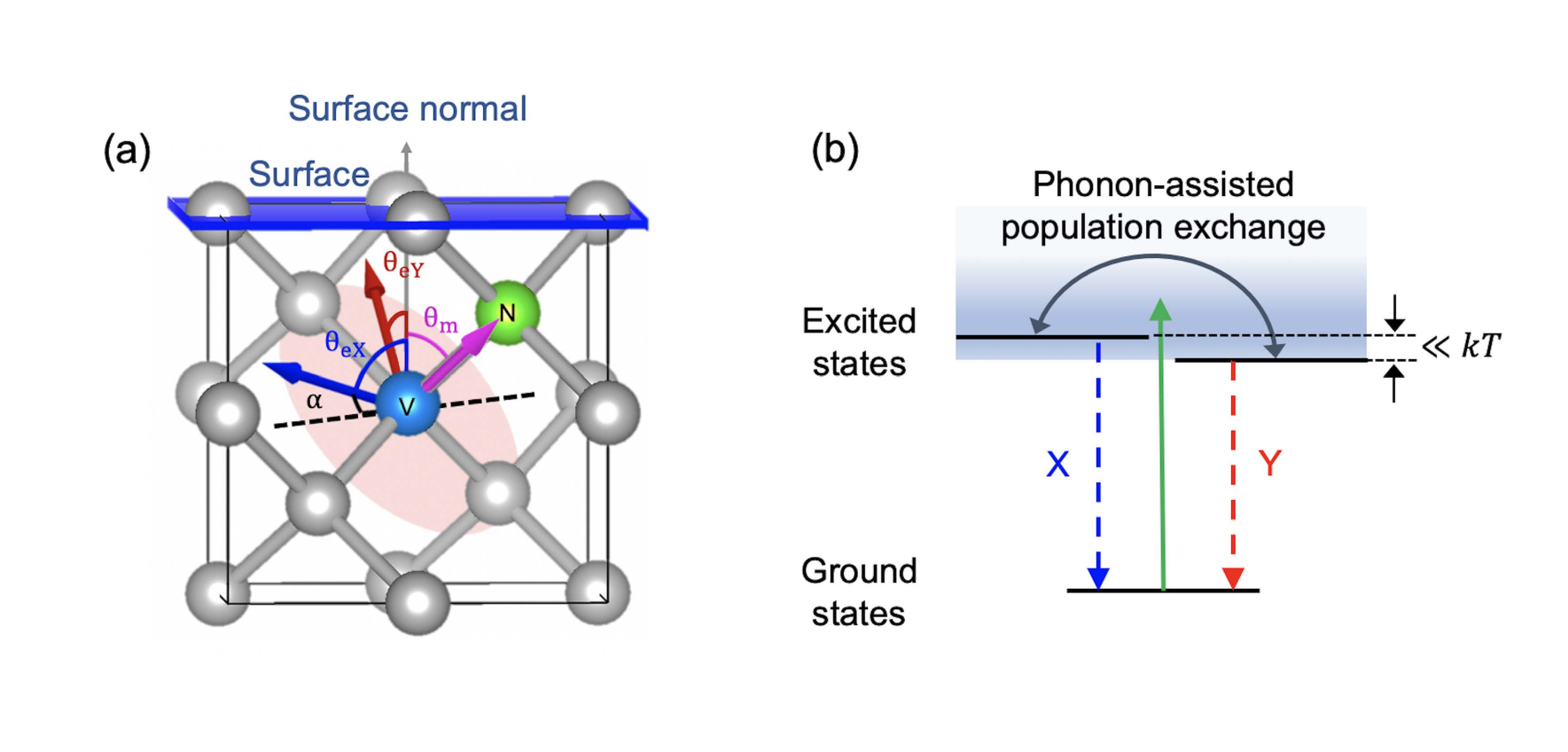}
    \caption{(a) Atomic structure for the NV center. $X$ and $Y$ are electric dipoles in the pink-colored plane perpendicular to the spin axis along [111] in (100)-oriented diamond. 
    If we project the surface plane to the location of the NV center, then it intersects with a plane that is perpendicular to the spin axis which
    is shown as a dashed line. \(\alpha\) is defined as an angle rotating clockwise from the intersection line to $X$, when observing from the opposite direction of the NV spin. Various angles are defined relative to the surface normal, to denote the orientations of the spin axis $\theta_m$ and the $X$ ($\theta_{eX}$) and Y ($\theta_{eY}$) dipoles. (b) Simplified energy level diagram for vacancy defects indicating the X and Y optical transitions. The transition energies for $X$ and $Y$ are shown to be offset by a small amount (much smaller than the thermal energy at room temperature) due to local strain.}
    \label{atomicStructure}
\end{figure}

Vacancy defects in diamond, including NV centers or split vacancy defects (such as most Group IV color centers) are formed from missing carbon atoms next to substitutional or interstitial atoms. The electron spin, whose interactions with external electromagnetic and strain fields provides the physical basis for most quantum sensing schemes, is oriented along one of the crystallographic axes ($i.e.$, $[111]$, $[1\overline{1}\overline{1}]$, $[\overline{1}1\overline{1}]$, or $[\overline{1}\overline{1}1]$). Optical emission from the spin defect arises from pairs of orthogonal electric dipoles (conventionally denoted as ${X}$ and ${Y}$) that are orthogonal to the spin axis \cite{hughes1967uniaxial,epstein2005anisotropic} (Figure~\ref{atomicStructure}a). The particular orientation for each set of electric dipoles for a given NV center 
is set by the non-axial local strain, such that for a large ensemble of emitters, the dipole orientations are homogeneously distributed on the plane perpendicular to the spin axis~\cite{radko2016determining}. It is worth noting that the dipole orientations can be preferentially aligned by mechanical strain, as shown by polarization-selective measurements of single NV centers in mechanically driven cantilevers \mbox{\cite{lee2016strain}}. 

This physical picture of the electric dipoles can be most clearly illustrated by polarization studies of NV excitation and photoluminescence in (111)-cut diamond at room temperature~\cite{alegre2007polarization, abe2017dynamically,rohner2019111} and near 4 K~\cite{batalov2009low,kaiser2009polarization, fu2009observation}. For an [111]-oriented NV in (111)-cut diamond, the spin axis is normal to the diamond surface while the electric dipoles are in the diamond surface plane. At low temperatures, resonant excitation of each of the ${X}$ and ${Y}$ dipoles with polarized light follows Malus's law, with the sinusoidal modulations from the two orbital branches exhibiting opposite phases~\cite{kaiser2009polarization}. The ability to distinguish between the polarized emissions from the $X$ and $Y$ transitions has a strong temperature dependence. At temperatures above 40 K, the emission from the [111]-oriented NV appears unpolarized \cite{abe2017dynamically,fu2009observation}.

For the NV center, the optical dipole transitions correspond to energies around 1.945 eV (or 637 nm in wavelength)~\cite{davies1976optical}. In fluorescence spectroscopy, the optical dipole transitions are observed as a zero phonon line followed by a broad phonon sideband that extend to $\sim800$ nm~\cite{jelezko2006single}. The branching ratio of the zero phonon line is typically $\sim3\%$~\cite{faraon2011resonant}. Local strain lifts the degeneracy of the two dipole moments, so that they are typically separated by a few GHz which is considerably below the thermal energy in most measurement scenarios (4 K to room temperature). Therefore, excitation of either dipole moment is followed by phonon-assisted population exchange between the excited states and eventually radiative decay to the ground states via one of the dipole transitions (Figure~\ref{atomicStructure}b). If the population exchange within the excited states occurs more quickly than radiative decay, then the excited-state populations achieve thermal equilibrium prior to decay leading to both dipole transitions occurring with equal probability~\cite{kaiser2009polarization} (see Supplementary Section 1). Since there is no correlation between subsequent emitted photons from $X$ and $Y$ electric dipoles, the emission is dynamically random and has been demonstrated experimentally in \mbox{\cite{abe2017dynamically}}. Therefore, we can assume that the unpolarized emission from the electric dipoles incoherently sums to yield radiation of photons that spectrally overlap at room temperature, with the overall rate of single-photon emission being the average between the radiative rate of each dipole moment ($R_X$ and $R_Y$): $R=(R_X + R_Y)/2$. This emission rate is proportional to the total radiative power, $L$, and is the inverse of the mean radiative lifetime $\tau_{mean}=1/R$.

In homogeneous dielectric environments (such as the case of an emitter deeply embedded in a bulk diamond crystal or an emitter-containing nanocrystal suspended in a uniform medium), $R_X = R_Y$ and therefore the mean lifetime is identical to the lifetime associated with each dipole transition (defined as $\tau_X$ and $\tau_Y$). In bulk diamond, this mean lifetime is typically $\sim13$ ns \cite{martin1999fine}. For an emitter in a non-isotropic medium ($e.g.$, near a surface or in nanostructures), the radiative rates from the $X$ and $Y$ dipoles can be affected unequally and thus contribute differently to the mean lifetime. The mean lifetime can then be calculated as a weighted average of the lifetimes (see Supplementary Section 1):

\begin{equation}
    \tau_{mean}=\frac{1}{R}=\frac{2}{R_X+R_Y}=\frac{R_X}{R_X+R_Y}\frac{1}{R_X}+\frac{R_Y}{R_X+R_Y}\frac{1}{R_Y}=\frac{R_X}{R_X+R_Y}\tau_X+\frac{R_Y}{R_X+R_Y}\tau_Y
    \label {weighted average lifetime}
\end{equation}

Finally, the total decay rate from an excited NV center is a sum of the rates from all possible radiative and non-radiative channels. The quantum efficiency quantifies the relative contributions of the radiative and non-radiative rates and is defined as the ratio between the radiative rate and the total decay rate. While only radiative transitions are shown in the simplified energy scheme in Figure~\ref{atomicStructure}b, excited NVs can also decay via non-radiative channels which include an intersystem crossing mechanism \mbox{\cite{goldman2015state}}, electron tunneling between NV centers and nitrogen impurities \mbox{\cite{capelli2022proximal}}, and recombination with other electron traps in the diamond crystal and at the interface \mbox{\cite{inam2014effects}}. For the NV center, the spin-dependent intersystem crossing decay is the basis of optical initialization and readout of the NV spin and occurs on a timescale much longer (on the order of hundreds of ns) than the radiative lifetime \cite{jelezko2006single}. Both the intersystem crossing and electron tunneling to nitrogen impurities are non-radiative processes that affect NVs regardless of their proximity to the surface. Meanwhile, non-radiative decays related to presence of electron acceptors on the diamond surface will be more prominent for near-surface NVs.

Since the quantum efficiency depends on the environment of each NV and can only be precisely determined experimentally \mbox{\cite{radko2016determining}}, our numerical analysis examines only the effect of NV depth on radiative transition rates. We will discuss the implications of the numerical results and impact of non-radiative processes on lifetime measurements in Section 4.


\section{Calculation of the radiative lifetime for an NV close to an interface}

\begin{figure}[ht]
    \centering
     \includegraphics[width=\textwidth]{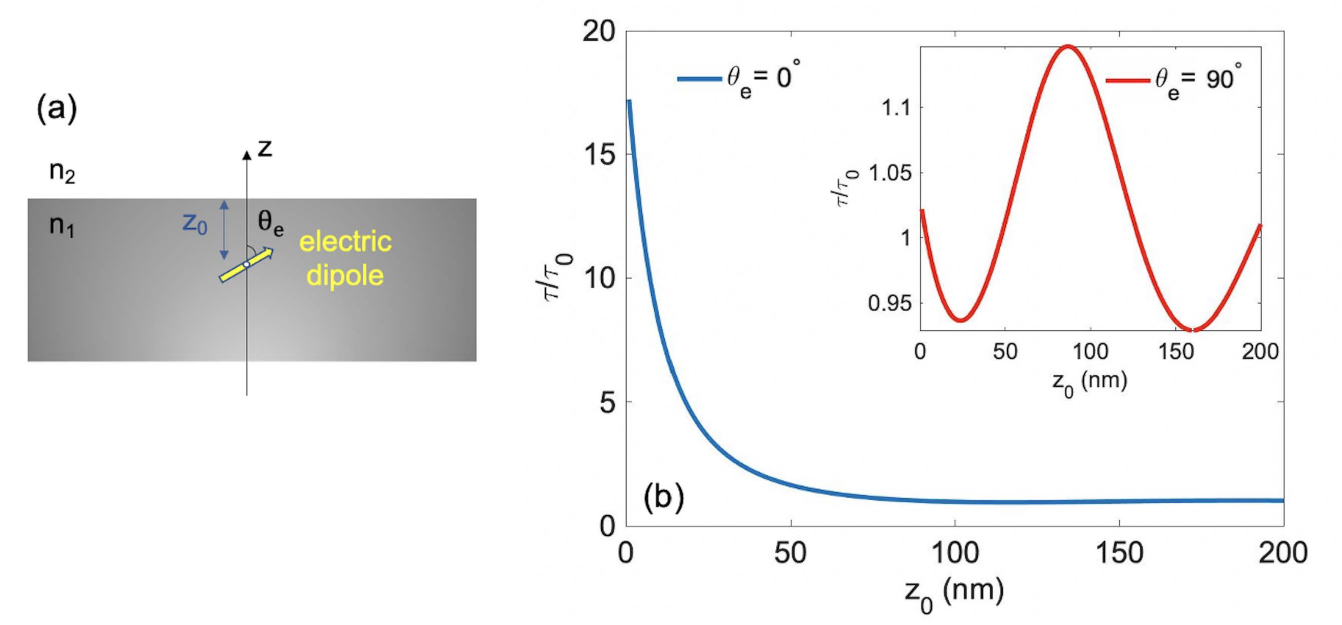}
    \caption{(a) Schematic showing an electric dipole at any angle ($\theta_{e}$) at distance $z_{0}$ from the interface. (b) Normalized lifetime of an electric dipole perpendicular (blue; $\theta_e=0^\circ$) and parallel (red; $\theta_e=90^\circ$) to the interface as a function of $z_{0}$ at $\lambda=637$ nm (zero phonon line of $NV^{-}$ in diamond), with $n_{1}=2.4$ (refractive index of diamond), and $n_{2}=1$ (refractive index of air).}
    \label{electric-dipoles}
\end{figure}

We use a numerical implementation of the angular spectrum method to calculate the radiative lifetimes of the $X$ and $Y$ dipoles near the surface. Figure~\ref{electric-dipoles}a illustrates an electric dipole inside diamond (with refractive index $n_1\sim2.4$) interfacing with air ($n_2\sim1$). The $z$ axis (which is also the surface cut of the diamond) is defined as the normal vector of the diamond-air interface while 
the depth $z_0$ is defined as the distance between the color center and the interface. The angle between the surface normal and the electric dipole is defined as $\theta_{e}$. Here, we use the subscript $e$ to denote electric dipole emission.

Since the lifetime of any electric dipole is inversely proportional to its radiative power~\cite{lukosz1977fluorescence}, we can calculate the depth-dependent lifetime ($\tau(z_0)$) of the dipole relative to its lifetime in homogeneous medium ($\tau(z_0)$) as follows:
\begin{equation}
  \tau(z_0)/\tau_{0}=[L(z_0)/L_{\infty}]_{e}^{-1}
  \label{lifetime}
\end{equation}
\noindent
where 
\([L({z_0})]_e\) and \([L_{\infty}]_e\) are the powers radiated by the electric dipole at \(z_0\) and \(z_0 \rightarrow \infty\). 

For any electric dipole at $\theta_{e}$, the radiative power ratio is comprised of angle-dependent contributions from the parallel ($[L(z_0)/L_{\infty}]_{e\parallel}$) and perpendicular ($[L(z_0)/L_{\infty}]_{e\perp}$) components of the radiative power ~\cite{lukosz1977fluorescence}:
\begin{equation} \label{general power ratio for single electric dipole}
  [L(z_0)/L_{\infty}]_{e}=\cos^2\theta_{e}[L(z_0)/L_{\infty}]_{e\perp}+\sin^2\theta_{e}[L(z_0)/L_{\infty}]_{e\parallel}.
\end{equation}
\noindent

\noindent
We adapt the equations introduced by Lukosz and Kunz \cite{lukosz1977fluorescence} to find the radiative power for an electric dipole, perpendicular and parallel to the surface as follows:

\begin{equation} 
  [L(z_0)/L_{\infty}]_{e\perp}=1+\frac{3}{2}\mathrm{Re}[(\int_0^1+\int_{i\infty}^0)r_{1,2}^{(p)}(1-\nu^2)\mathrm{exp}{(2i k^{(1)}z_0\nu)}\ d\nu]
 \label{perp power ratio}
\end{equation}

\begin{equation} 
  [L(z_0)/L_{\infty}]_{e\parallel}=\frac{1}{2}\{1+[L(z_0)/L_{\infty}]_{e,\perp}\}
+\frac{3}{4}\mathrm{Re}\{(\int_0^1+\int_{i\infty}^0)[r_{1,2}^{(s)}-r_{1,2}^{(p)}]\mathrm{exp}{(2i k^{(1)}z_0\nu)}\ d\nu\}
\label{para power ratio}
\end{equation}

\noindent
where $k^{(1)}$ and $k^{(2)}$ are respectively wave vectors in medium 1 and 2; $\nu={k_z^{(1)}}/{k^{(1)}}$ is defined as the normalized z-component of the wave vector in medium 1. $r_{1,2}^{(s)}$ and $r_{1,2}^{(p)}$ are reflection coefficients for s and p polarized light (see Supplementary Section 3):

\noindent

\begin{equation} \label{rs-nu}
  r_{1,2}^{(s)}(\nu)=\frac{k^{(1)}\nu-k^{(2)}\sqrt{1-\frac{1}{n^2}+\frac{1}{n^2}\nu^2}}{k^{(1)}\nu+k^{(2)}\sqrt{1-\frac{1}{n^2}+\frac{1}{n^2}\nu^2}}
\end{equation}

\begin{equation}\label{rp-nu}
  r_{1,2}^{(p)}(\nu)=\frac{\frac{\epsilon_2}{\epsilon_1}k^{(1)}\nu-k^{(2)}\sqrt{1-\frac{1}{n^2}+\frac{1}{n^2}\nu^2}}{\frac{\epsilon_2}{\epsilon_1}k^{(1)}\nu+k^{(2)}\sqrt{1-\frac{1}{n^2}+\frac{1}{n^2}\nu^2}}
\end{equation}

\noindent where $\epsilon_{1}$ and $\epsilon_{2}$ are the dielectric constants in medium 1 and 2 and $n={n_2}/{n_1}$. To acquire the lifetime shown in Equation~\ref{lifetime}, the radiative power ratio was calculated by numerically solving the integrals of Equations \ref{perp power ratio} and \ref{para power ratio}. We note that the terms involving integration over the imaginary plane (Re$\int_{i\infty}^0$) are zero except when $n>1$ ($e.g.$ diamond is in contact with a higher index material), due to the presence of evanescent waves at the interface \cite{lukosz1977fluorescence}. Please note that Equations \mbox{\ref{perp power ratio}} and \mbox{\ref{para power ratio}} are only valid in case of non-absorbing media where \mbox{$n_1$} and \mbox{$n_2$} are real.
 
The depth-dependent lifetimes are shown in Figure \ref{electric-dipoles}b and exhibit monotonic increase (up to $>16$-fold at the interface) for the perpendicular electric dipole at $z_{0}<200$ nm, while the radiative lifetime of the parallel electric dipole modulates slightly (with $<15\%$ variation) with depth. These results are consistent with previous studies of classical dipoles near interfaces \cite{drexhage1970influence, cui2015reduced} and have their physical origins in the constructive and destructive interference between the dipole emissions and the reflected waves from the interface. In addition, we calculated the changes in lifetime of the electric dipole close to an interface with $n>1$ and observed a decrease in lifetime as expected (see Supplementary Section 5).

To account for the strong presence of the phonon sideband in the NV emission described in Section \ref{photo-physics}, the spectrally averaged powers (represented by $\langle[L(z_0)/L_{\infty}]_{e \perp}\rangle$ and $\langle[L(z_0)/L_{\infty}]_{e \parallel}\rangle$ for electric dipoles perpendicular and parallel to the interface) are calculated as follows:
 
 \begin{equation}
    \langle[L(z_0)/L_{\infty}]_{e \perp/\parallel}\rangle = \frac{ \int [L(z_0)/L_{\infty}]_{e \perp/\parallel}I_{NV}(\lambda)d\lambda}{\int I_{NV}(\lambda)d\lambda} 
\end{equation}

\noindent
where $I_{NV}(\lambda)$ is the emission spectrum of NV acquired at room temperature for an NV in bulk diamond.

Using Equations~\ref{lifetime} and ~\ref{general power ratio for single electric dipole}, we can now calculate the lifetime of each of the $X$ and $Y$ dipoles of the NV: 

\begin{equation}
  [\tau(z_0,\alpha,\theta_m)/\tau_0]_{X/Y}=[\cos^2\theta_{eX/Y}\langle[L(z_0)/L_{\infty}]_{e\perp}\rangle+\sin^2\theta_{eX/Y}\langle[L(z_0)/L_{\infty}]_{e\parallel}\rangle]^{-1}
  \label{lifetime-X}
\end{equation}

\noindent
where $\theta_{eX}$ and $\theta_{eY}$ are the angles between the surface normal and the $X$ and $Y$ electric dipoles, with
\begin{align}
X:
    \begin{cases}
    \sin\theta_{eX}=\sin(\alpha)\sin(\theta_m)\\
    \cos\theta_{eX}=\sqrt{1-\sin^2(\alpha)\sin^2(\theta_m)}
    \end{cases} 
    &
    Y:
    \begin{cases}
    \sin\theta_{eY}=\cos(\alpha)\sin(\theta_m)\\
    \cos\theta_{eY}=\sqrt{1-\cos^2(\alpha)\sin^2(\theta_m)}
    \end{cases} 
\label{theta_x}
\end{align}
   
Equations~\ref{lifetime-X} and \ref{theta_x} demonstrate the dependence of $\tau_X$ and $\tau_Y$ on the spin axis orientation ($\theta_m$) and the orientation of the dipoles on the plane perpendicular to the spin axis (defined through angle $\alpha$ as shown in Figure~\ref{atomicStructure}a). These relationships can be combined with Equation~\ref {weighted average lifetime} to yield the mean radiative lifetime for a single NV (see Supplementary Section 4):

\begin{equation} \label{averaging power ratio for two electric dipoles of one color center}
\begin{split}
 [\tau(z_0,\theta_m)/\tau_{0}]_{mean}
 =[\frac{1}{2}\sin^2\theta_m\langle[L(z_0)/L_{\infty}]_{e\perp}\rangle+\frac{1}{2}(1+\cos^2\theta_m)\langle[L(z_0)/L_{\infty}]_{e\parallel}\rangle]^{-1}
\end{split}
\end{equation}


\noindent
It can be inferred from Equation~\ref{averaging power ratio for two electric dipoles of one color center} that the mean radiative lifetime for any NV is a function of \(\theta_m\) and $z_{0}$ but independent of \(\alpha\). As shown in Figures \mbox{\ref{100 surface cut}}(c), \mbox{\ref{110 surface cut}}(c), and \mbox{\ref{111 surface cut}}(c), the $\alpha$-dependent variations in $\tau_{X}$ and $\tau_{Y}$ are 90$^\circ$ out of phase with one another, so that the mean radiative lifetime is the same for all values for $\alpha$. Therefore, for any arbitrary pairs of orthogonal dipoles \(X\) and \(Y\) on the plane normal to the defined spin axis, the depth dependence of the mean radiative lifetime is the same.

\section{Results and Discussion}

We now explicitly calculate the NV radiative lifetimes associated with near-surface spins along the four spin quantization axes in (100)-, (110)-, and (111)-diamond. Table \ref{power ratio in different surface cut} shows the possible $\theta_m$ values for any NV and the corresponding spectrally averaged weights of $\langle[L(z_0)/L_{\infty}]_{e\perp}\rangle$ and $\langle[L(z_0)/L_{\infty}]_{e\parallel}\rangle$ in the determination of the total radiative power $\langle[L(z_0)/L_{\infty}]_e\rangle$. Here, we use the notation $NV_i$ ($i=1,2,3,4$) to respectively indicate the four possible spin axes in diamond $[111]$, $[1\overline{1}\overline{1}]$, $[\overline{1}1\overline{1}]$ and $[\overline{1}\overline{1}1]$ which can be determined by magnetometry and/or polarization-resolved measurements \cite{alegre2007polarization,schloss2018simultaneous, mccullian2022quantifying}. Due to trigonal symmetry, the normalized power of NVs in (100)-oriented diamond is the sum of the perpendicular and parallel components with respective weights $\frac{1}{3}$ and $\frac{2}{3}$, regardless of the NV spin orientation. This result is consistent with calculations shown in~\cite{ radtke2019nanoscale, radko2016determining}. For (110)-diamond, two unique power distributions are derived for NVs with $\theta_m=35.3^\circ, 144.7^\circ$ and $\theta_m = 90^\circ$. Since the radiative power of the perpendicular dipole component experiences more quenching as the emitter approaches the surface (Figure~\ref{electric-dipoles}b), we expect the NVs with $\theta_m = 90^\circ$ which have a larger contribution from $[{L_{\infty}}]_{e\perp}$, to be more sensitive to their proximity to the surface. Finally, NVs in (111)-diamond with the spin axis perpendicular to the surface radiate similarly to a parallel dipole and be minimally affected by depth, while the other three orientations (all with $\theta_m = 109.5^\circ$) show stronger depth dependence.

\begin{table}[H]

\centering
\renewcommand{\arraystretch}{2}
\scalebox{0.9}
{
\begin{tabular}{|c|c|c|c|c|}
\hline
NV direction& $NV_1:[111]$ & $NV_2:[1\overline{1}\overline{1}]$ & $NV_3:[\overline{1}1\overline{1}]$ &$NV_4:[\overline{1}\overline{1}1]$\\

\hline\hline
\multicolumn{5}{|c|}{(100) surface} \\
\hline
$\theta_m$ & $54.7^{\circ}$  &  $54.7^{\circ}$ & $125.3^{\circ}$  & $125.3^{\circ}$ \\
\hline
$\langle[\frac{L(z_0)}{L_{\infty}}]_e\rangle$ & $\frac{1}{3}\langle[\frac{L(z_0)}{L_{\infty}}]_{e\perp}\rangle+\frac{2}{3}\langle[\frac{L(z_0)}{L_{\infty}}]_{e\parallel}\rangle$  &  $\frac{1}{3}\langle[\frac{L(z_0)}{L_{\infty}}]_{e\perp}\rangle+\frac{2}{3}\langle[\frac{L(z_0)}{L_{\infty}}]_{e\parallel}\rangle$ & $\frac{1}{3}\langle[\frac{L(z_0)}{L_{\infty}}]_{e\perp}\rangle+\frac{2}{3}\langle[\frac{L(z_0)}{L_{\infty}}]_{e\parallel}\rangle$  & $\frac{1}{3}\langle[\frac{L(z_0)}{L_{\infty}}]_{e\perp}\rangle+\frac{2}{3}\langle[\frac{L(z_0)}{L_{\infty}}]_{e\parallel}\rangle$ \\

\hline\hline
\multicolumn{5}{|c|}{(110) surface} \\
\hline
$\theta_m$ & $35.3^{\circ}$  &  $90.0^{\circ}$ & $90.0^{\circ}$  & $144.7^{\circ}$ \\
\hline
$\langle[\frac{L(z_0)}{L_{\infty}}]_e\rangle$ & $\frac{1}{6}\langle[\frac{L(z_0)}{L_{\infty}}]_{e\perp}\rangle+\frac{5}{6}\langle[\frac{L(z_0)}{L_{\infty}}]_{e\parallel}\rangle$  &  $\frac{1}{2}\langle[\frac{L(z_0)}{L_{\infty}}]_{e\perp}\rangle+\frac{1}{2}\langle[\frac{L(z_0)}{L_{\infty}}]_{e\parallel}\rangle$ & $\frac{1}{2}\langle[\frac{L(z_0)}{L_{\infty}}]_{e\perp}\rangle+\frac{1}{2}\langle[\frac{L(z_0)}{L_{\infty}}]_{e\parallel}\rangle$  & $\frac{1}{6}\langle[\frac{L(z_0)}{L_{\infty}}]_{e\perp}\rangle+\frac{5}{6}\langle[\frac{L(z_0)}{L_{\infty}}]_{e\parallel}\rangle$ \\

\hline\hline
\multicolumn{5}{|c|}{(111) surface} \\
\hline
$\theta_m$ & $0.0^{\circ}$  &  $109.5^{\circ}$ & $109.5^{\circ}$  & $109.5^{\circ}$ \\
\hline
$\langle[\frac{L(z_0)}{L_{\infty}}]_e\rangle$ & $\langle[\frac{L(z_0)}{L_{\infty}}]_{e\parallel}\rangle$  &  $\frac{4}{9}\langle[\frac{L(z_0)}{L_{\infty}}]_{e\perp}\rangle+\frac{5}{9}\langle[\frac{L(z_0)}{L_{\infty}}]_{e\parallel}\rangle$ & $\frac{4}{9}\langle[\frac{L(z_0)}{L_{\infty}}]_{e\perp}\rangle+\frac{5}{9}\langle[\frac{L(z_0)}{L_{\infty}}]_{e\parallel}\rangle$  & $\frac{4}{9}\langle[\frac{L(z_0)}{L_{\infty}}]_{e\perp}\rangle+\frac{5}{9}\langle[\frac{L(z_0)}{L_{\infty}}]_{e\parallel}\rangle$ \\
\hline

\end{tabular}}
\caption{$\theta_m$ and power ratio for different NV directions in different surface cuts.}
\label{power ratio in different surface cut}
\end{table}

\begin{figure}[ht]
    \centering
    \includegraphics[width=\textwidth]{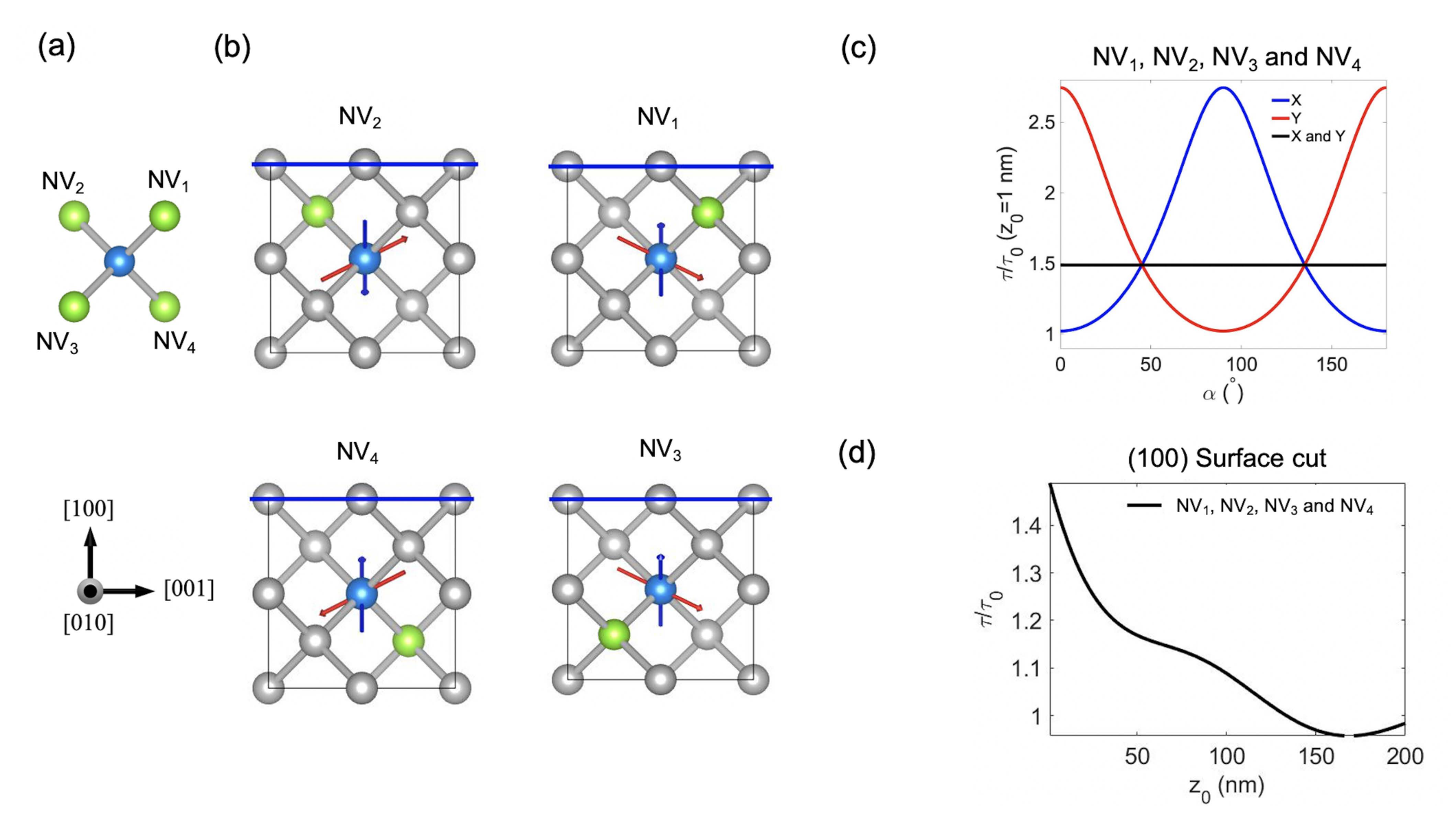}
    \caption{(a) 3D plot of NVs along four possible orientations of crystallographic axes in a (100)-oriented diamond and its coordinate axes. (b) Atomic structure of the NV centers and arbitrary set of $X$ (blue) and $Y$ (red) electric dipoles associated with each NV. (c) The radiative lifetime ratios for the $X$ (blue) and $Y$ (red) electric dipoles at $z_{0}$=1 nm, along with the mean lifetime (black), as a function of $\alpha$. (d) The mean radiative lifetime ratio for $NV_1$, $NV_2$, $NV_3$, and $NV_4$ as a function of $z_{0}$.}
    \label{100 surface cut}
\end{figure}

\begin{figure}[ht]
    \centering
    \includegraphics[width=\textwidth]{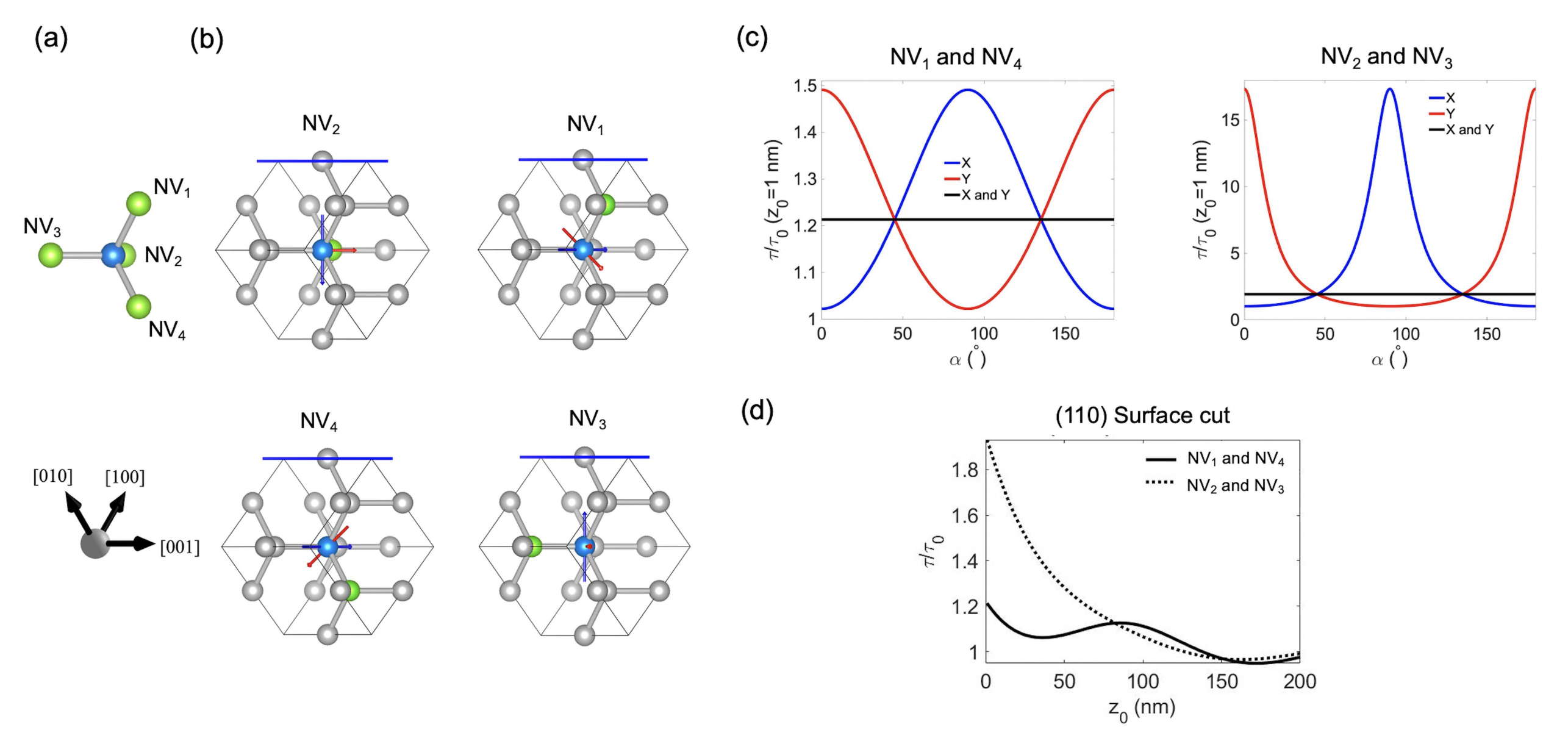}
    \caption{(a) 3D plot of NVs along four possible orientations of crystallographic axes in a (110)-oriented diamond and its coordinate axes. (b) Atomic structure of the NV centers and arbitrary set of $X$ (blue) and $Y$ (red) electric dipoles associated with each NV. (c) The radiative lifetime ratios for the $X$ (blue) and $Y$ (red) electric dipoles at $z_{0}$=1 nm, along with the mean lifetime (black), as a function of $\alpha$. (d) The mean radiative lifetime ratio for $NV_1$, $NV_2$, $NV_3$, and $NV_4$ as a function of $z_{0}$.}
    \label{110 surface cut}
\end{figure}

\begin{figure}[ht]
    \centering
    \includegraphics[width=\textwidth]{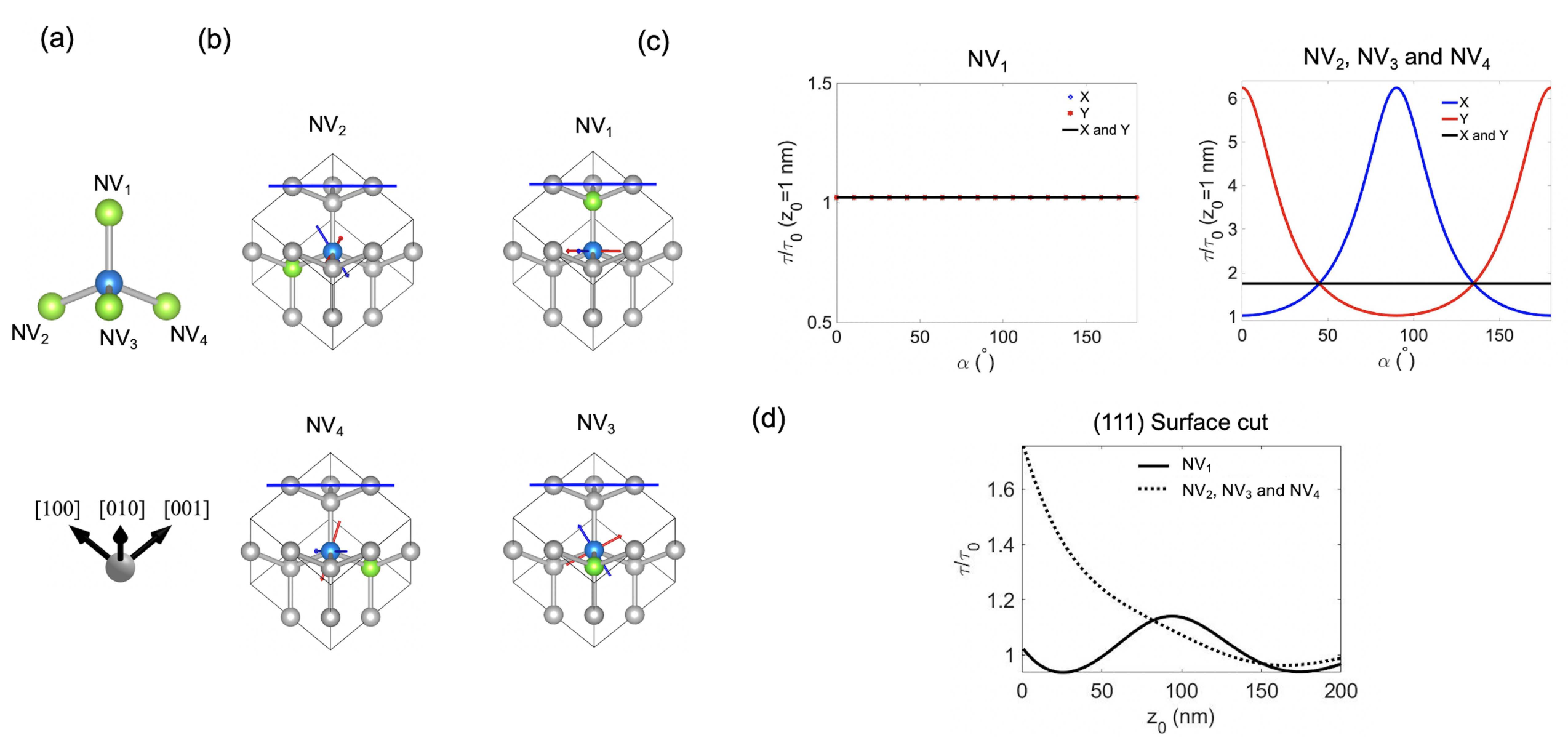}
    \caption{(a) 3D plot of NVs along four possible orientations of crystallographic axes in a (111)-oriented diamond and its coordinate axes. (b) Atomic structure of the NV centers and arbitrary set of $X$ (blue) and $Y$ (red) electric dipoles associated with each NV. (c) The radiative lifetime ratios for the $X$ (blue) and $Y$ (red) electric dipoles at $z_{0}$=1 nm, along with the mean lifetime (black), as a function of $\alpha$. (d) The mean radiative lifetime ratio for $NV_1$, $NV_2$, $NV_3$, and $NV_4$ as a function of $z_{0}$.}
    \label{111 surface cut}
\end{figure}

The radiative lifetime in each case is then calculated as an inverse of the radiated power. Our depth-dependent lifetimes for all possible NV orientations are plotted in Figures~\ref{100 surface cut},~\ref{110 surface cut}, and~\ref{111 surface cut} for (100)-, (110)-, and (111)-diamond. For each possible NV orientation, we also show the variation in the lifetimes of individual dipole transitions $X$ and $Y$ as a function of $\alpha$ at $z_0=1$ nm using Equation~\ref{lifetime-X}, along with the mean lifetime of the NV emission which is invariant to $\alpha$ (Figures~\ref{100 surface cut}c,~\ref{110 surface cut}c, and~\ref{111 surface cut}c).

Figures~\ref{100 surface cut}d,~\ref{110 surface cut}d, and~\ref{111 surface cut}d show that proximity to the surface can significantly modify the lifetime at depths shallower than $\sim$ 100 nm, but the effect greatly depends on the NV orientation and surface cut. In (100)-diamond, there is a monotonic increase in the mean radiative lifetime as NVs are closer to the surface, leading to a $\sim$1.5-fold increase in the lifetime at the surface. For defects in (110)-diamond, the mean radiative lifetime for NVs with $\theta_m=90^{\circ}$ increases by a factor of $> 1.8$ within a few nanometers from the surface. Meanwhile, NVs with $\theta_m=35.3^{\circ}$ and $144.7^\circ$ experience small increase as well as sinusoidal modulation in the lifetime with decreasing depth, due to self-interference between the emitter and reflected light. Similarly for $NV_1$ with $\theta_m=0^{\circ}$ in (111)-diamond, the mean lifetime is slightly modulated by depth (Figure \ref{111 surface cut}d). The other three NV centers with identical $\theta_m=109.5^{\circ}$ in (111)-oriented diamond, exhibit $\sim$1.8-fold increase in their mean radiative lifetime (Figure \ref{111 surface cut}d).

We also performed the lifetime calculation for an NV in diamond interfacing with different dielectrics (see Supplementary Section 5). As expected, a smaller mismatch in the dielectric constants reduces the suppression in radiative power, resulting in a reduction of $7\%$ in the lifetime increase between a diamond-air and diamond-oil interface. Meanwhile NVs close to an interface with higher refractive index will exhibit enhanced emission rate and corresponding shortening in the lifetime, with $>35\%$ lifetime reduction in comparison to bulk for a silicon-like material.

Accurate determination of the color center depth is important for nuclear magnetic resonance (NMR) spectroscopy using spin defects \cite{devience2015nanoscale} as well as correlation studies between spin coherence and proximity to sources of electric and magnetic noise \cite{sangtawesin2019origins}. Current methods for depth determination tend to be destructive ($e.g.$ Secondary ion mass spectroscopy \cite{toyli2010chip, fiori2014improved}) or rely on interfacing diamond with another material that provides sources of electron acceptors (in the case of FRET \cite{tisler2013single}) or proton spins (for NMR techniques \cite{pham2016nmr}). We thus consider the prospect of using excited-state-lifetime measurement as a simple technique to estimate emitter depth in diamond.

While our numerical analysis considered only the depth-dependence of the radiative lifetime, in measurements non-radiative processes reduce both the fluorescence intensity and the excited state lifetime. Knowledge of the emitter's quantum efficiency is thus needed to infer the radiative lifetime from measurements. Radko et al. experimentally determined the internal quantum efficiency for NVs in bulk diamond to be between $\sim$0.7 and $\sim$0.86, with the lower quantum efficiency values associated with NVs a few nm from the surface \mbox{\cite{radko2016determining}}, suggesting that non-radiative processes are slightly enhanced near the surface. In addition to non-radiative decays, there may be excitation and decay of surface defects. However, the observed lifetimes associated with these defects tend to be fast (few ns) and thus can be discerned from NV emission by multi-exponential fitting \mbox{\cite{tisler2013single, smith2009five,inam2013emission, mohtashami2013suitability}}.

Due to the aforementioned complications along with the moderate dynamic range of the lifetime variation, exact comparison between our numerical model with experiments can be challenging. Nonetheless, our numerical results for (100)-surface cut diamond are consistent with experimental data from Radtke et al and Radkko et al \mbox{\cite{radtke2019nanoscale, radko2016determining}}, which demonstrated that shallow NVs $\lesssim$ 10 nm from the surface have prolonged lifetimes $\sim 1.2$ to 1.3 $\times$ NV lifetimes deep in the bulk.  
To estimate the emitter-dependent variation in quantum efficiency, it may be helpful to correlate the relative difference in lifetimes with the relative difference in fluorescence intensity (e.g., as measured at saturation) \mbox{\cite{choy2011enhanced}}. This would allow for qualitative comparison of emitter depth within the same diamond sample.

Finally, the calculations presented here are applicable for other color centers in diamond such as Group-IV-split-vacancy centers~\cite{bradac2019quantum}, where polarization-selective photoluminescence studies revealed the presence of sets of electric dipoles similar to NV centers \cite{hepp2014electronic}.

\medskip
\textbf{Supporting Information} \par 
Supporting Information can be found towards the end of the document.

\medskip
\textbf{Acknowledgements} \par 
This work is supported by the U.S. Department of Energy, Office of Science, Basic Energy Sciences under
Award $\#$DE-SC0020313. We are grateful for insightful discussions with Mikhail Kats on modeling of the NV dipoles.

\medskip
\newpage
\section*{References}
\bibliographystyle{apsrev4-1}
\bibliography{Bibliography}

\newpage
\section*{Supplementary information}
\subsection{Rate equations for single-photon emission from an NV} \label{photo-physics-decay}
\renewcommand{\thepage}{S\arabic{page}}  
\renewcommand{\thesection}{S\arabic{section}}   
\renewcommand{\thetable}{S\arabic{table}}   
\renewcommand{\thefigure}{S\arabic{figure}}

As described in the main text, optical emission from the NV arises from two orthogonal electric dipoles denoted as $X$ and $Y$. While each of the $X$ and $Y$ dipole moments can be independently excited, phonon processes lead to population exchange between the excited states. Eventually, radiative decay can occur via either transition. The rate equations describing the radiative decay of an excited NV can be written as:

\begin{equation}
  \begin{cases} 
    \frac{dP_X}{dt}=-R_XP_X-\gamma(P_X-P_Y)\\
    \frac{dP_Y}{dt}=-R_YP_Y-\gamma(P_Y-P_X)
  \end{cases}
\end{equation}

\noindent where \(R_X\) and \(R_Y\) are the decay rates from states \(X\) and \(Y\); $\gamma$ is the rate of phonon-assisted population exchange between the \(X\) and \(Y\) excited states; \(P_X\) and \(P_Y\) are the normalized excited state populations at \(X\) and \(Y\). We note that non-radiative decay processes of the NV through the singlet states can be neglected since they occur on much longer timescales. Solving the set of rate equations yields:

\begin{align*}
&\Rightarrow
    \frac{d^2P_X}{dt^2}+(R_X+R_Y+2\gamma)\frac{dP_X}{dt}+(R_XR_Y+R_X\gamma+R_Y\gamma)P_X=0\\
&\Rightarrow
    P_X=A e^{\beta_1 t}+B e^{\beta_2 t}, \mathrm{where}
\end{align*}
\begin{equation}
\begin{cases}
  \beta_1=\frac{-(R_X+R_Y+2\gamma)+\sqrt{(R_X-R_Y)^2+4\gamma^2}}{2}\\
  \beta_2=\frac{-(R_X+R_Y+2\gamma)-\sqrt{(R_X-R_Y)^2+4\gamma^2}}{2}
\end{cases}
\end{equation}

\noindent
We set the initial populations to be \(P_X(0)=P_Y(0)=\frac{1}{2}\), corresponding to the case in which both $X$ and $Y$ dipole moments have equal probabilities of being excited. The time-dependent excited state populations are then:
\begin{equation}
    P_X=\frac{\beta_2+R_X}{2(\beta_2-\beta_1)}e^{\beta_1 t}
    +\frac{\beta_1+R_X}{2(\beta_1-\beta_2)}e^{\beta_2 t}\\
\end{equation}


\begin{equation}
 P_Y=\frac{\beta_2+R_Y}{2(\beta_2-\beta_1)}e^{\beta_1 t}
    +\frac{\beta_1+R_Y}{2(\beta_1-\beta_2)}e^{\beta_2 t}
\end{equation}
\noindent The total excited state population, $P$, is the sum of $P_X$ and $P_Y$ and is proportional to the number of emitted photons after excitation in a fluorescence measurement. Using $\tau_X = \tau_0$ and $\tau_Y = 2.7 \tau_0$ (a valid set of orthogonal dipoles for (100)-diamond as shown in Figure 3c in the main text), we plot the time trace of excited state population for different values of $\gamma$ in Figure~\ref{rate_eqn_soln_100}. Our results show that for for $\gamma^{-1}\ll\tau_0$, the mean lifetime corresponds to the inverse of the average emission rates between $R_X$ and $R_Y$, as follows:

\begin{figure}[t]
    \centering
    \includegraphics[width=\textwidth]{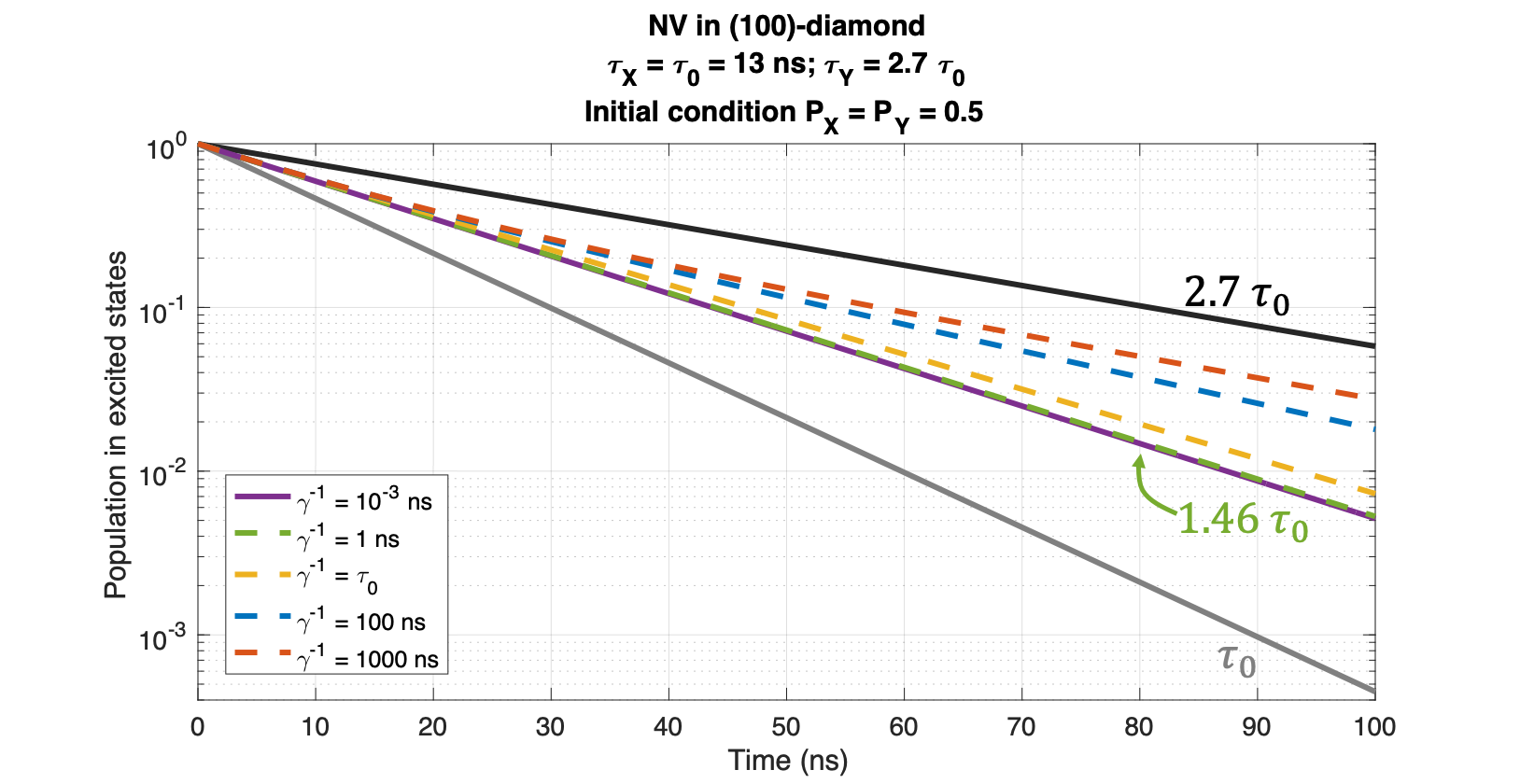}
    \caption{Time trace of excited state population for an NV in (100)-diamond, for different $\gamma$ values.}
    \label{rate_eqn_soln_100}
\end{figure}

\begin{align*}
&\Rightarrow
  \begin{cases}
    \beta_1\approx-\frac{R_X+R_Y}{2}\\
    \beta_2\approx-\frac{R_X+R_Y+4\gamma}{2}
  \end{cases}\\
&\Rightarrow
  \begin{cases}
    P_X\approx\frac{1}{2}e^{-(R_X+R_Y)t/2}\\
    P_Y\approx\frac{1}{2}e^{-(R_X+R_Y)t/2}
  \end{cases}
\end{align*}
\noindent
So the total radiated rate $R$ for the whole system becomes:
\begin{equation}
\label{total rate}
    R=\frac{R_X+R_Y}{2}.
\end{equation}
\noindent
$\gamma$ is dependent on the temperature and is fast ($\gg\tau_0^{-1}$) at room temperature~\cite{martin1999fine, fu2009observation}, leading to the appearance of unpolarized emission~\cite{kaiser2009polarization}. As $\gamma$ decreases with temperature, polarized emissions from the $X$ and $Y$ transitions can be distinguished. In the regime that $\gamma\lesssim\tau_0^{-1}$, the excited state decay tends towards a bi-exponential behavior with longer time constants.

\subsection{Determination of orthogonal dipole orientations} 

In this section, we describe our approach to generating valid sets of dipole orientations for our lifetime calculations. As noted in the main text, a set of orthogonal electric dipoles for a given NV center can be oriented along any direction on the plane perpendicular to the spin axis and thus the solutions provided here are not unique. 

If we define $X$ and $Y$ dipole directions for $NV_1$ to be $X\parallel[\overline{1}\overline{1}2]$ and $Y\parallel[1\overline{1}0]$, then we use rotation matrix rotating from $NV_1$ to other three different direction to find different $X$ and $Y$ for different $NV$ direction.

If we want to find the rotation matrix $A$ that rotates $\hat{NV_1}$ to $\hat{NV_2}$, we need to know 
\begin{align}
    &\vec{w}=\hat{NV_1}\times\hat{NV_2}=\frac{2}{3}[01\overline{1}]\\
    &s=\lVert w\lVert \textrm{ (sine of angle)}\\
    &c=\hat{NV_1}\cdot\hat{NV_2}  \textrm{ (cosine of angle)}
\end{align}

Then the rotation matrix $A$ is 
\begin{equation}
    A=I+[w]_\times+\frac{1-c}{s^2}[w]_\times^2
\end{equation}

\noindent
where $[w]_\times$ is the skew-symmetric cross-product matrix of $w$

\begin{equation}
    [w]_\times\stackrel{\text{def}}{=} \begin{bmatrix} 
	0 & -w_3 & w_2 \\
	w_3 & 0 & -w_1\\
	-w_2 & w_1 & 0 \\
	\end{bmatrix}=\frac{2}{3}\begin{bmatrix} 
	0 & 1 & 1 \\
	-1 & 0 & 0\\
	-1 & 0 & 0 \\
	\end{bmatrix}
\end{equation}

Then we can get rotation matrix $A= \frac{1}{3}\begin{bmatrix} 
	-1 & 2 & 2 \\
	-2 & 1 & -2\\
	-2 & -2 & 1 \\
	\end{bmatrix}$

We apply this rotation matrix to $X$ and $Y$, then we can get $X^*$ and $Y^*$ for $NV_2$
\begin{align}
    X^*&=AX=\frac{1}{3}\begin{bmatrix} 
	-1 & 2 & 2 \\
	-2 & 1 & -2\\
	-2 & -2 & 1 \\
	\end{bmatrix}
	\begin{bmatrix} 
	-1 \\
	-1 \\
	2  \\
	\end{bmatrix}=
	\begin{bmatrix} 
	1 \\
	-1 \\
	2  \\
	\end{bmatrix}\\
	Y^*&=AY=\frac{1}{3}\begin{bmatrix} 
	-1 & 2 & 2 \\
	-2 & 1 & -2\\
	-2 & -2 & 1 \\
	\end{bmatrix}
	\begin{bmatrix} 
	1 \\
	-1 \\
	0  \\
	\end{bmatrix}=
	\begin{bmatrix} 
	-1 \\
	-1 \\
	0  \\
	\end{bmatrix}
\end{align}

Using the same method, we can find out the direction of $X$ and $Y$ for three other $NV$ directions shown in table~\ref{4 directions of X and Y}.

\begin{table}[ht]
\centering
\renewcommand{\arraystretch}{1.5}
\begin{tabular}{ |c|c|c|c|c|}
\hline
NV direction& $NV_1:[111]$ & $NV_2:[1\overline{1}\overline{1}]$ & $NV_3:[\overline{1}1\overline{1}]$ &$NV_4:[\overline{1}\overline{1}1]$\\

\hline\hline
X direction& $[\overline{1}\overline{1}2]$  &  $  [1\overline{1}2]$ & $[\overline{1}12]$  & $[\overline{1}\overline{1}\overline{2}]$ \\
\hline
Y direction& $[1\overline{1}0]$ & $[\overline{1}\overline{1}0]$ & $[110]$ & $[1\overline{1}0]$\\

\hline\hline
\multicolumn{5}{|c|}{$\theta_e$ angle for (100) surface cut } \\
\hline
$\theta_{eX}$ & $114.1^{\circ}$  &  $65.9^{\circ}$ & $114.1^{\circ}$  & $114.1^{\circ}$ \\
\hline
$\theta_{eY}$ & $45.0^{\circ}$  &  $135.0^{\circ}$ & $45.0^{\circ}$  & $45.0^{\circ}$ \\

\hline\hline
\multicolumn{5}{|c|}{$\theta_e$ angle for (110) surface cut } \\
\hline
$\theta_{eX}$ & $125.3^{\circ}$  &  $90.0^{\circ}$ & $90.0^{\circ}$  & $125.3^{\circ}$ \\
\hline
$\theta_{eY}$ & $90.0^{\circ}$  &  $180.0^{\circ}$ & $0.0^{\circ}$  & $90.0^{\circ}$ \\

\hline\hline
\multicolumn{5}{|c|}{$\theta_e$ angle for (111) surface cut } \\
\hline
$\theta_{eX}$ & $90.0^{\circ}$  &  $61.9^{\circ}$ & $61.9^{\circ}$  & $160.5^{\circ}$ \\
\hline
$\theta_{eY}$ & $90.0^{\circ}$  &  $144.7^{\circ}$ & $35.3^{\circ}$  & $90.0^{\circ}$ \\
\hline

\end{tabular}
\caption{X and Y dipole directions for four different NV directions and the angle between the normal vector of surface cut and X or Y}
\label{4 directions of X and Y}
\end{table}

\subsection{Comparison of numerical integration with full-wave simulation and Taylor expansion approaches} \label{Theory_Compare}

To calculate $[L(z_0)/L_{\infty}]_{e,\perp}$ and $[L(z_0)/L_{\infty}]_{e,\parallel}$, we explored using the Finite-Difference Time-Domain (FDTD) method \cite{taflove2005computational} (using commercial software Lumerical) in addition to numerically integrating Equations. 4 and 5.

To solve the integrals, we need to first find $r_{1,2}^{(s)}(\nu)$ and $r_{1,2}^{(p)}(\nu)$, where reflection coefficients are defined as follows~\cite{lukosz1977fluorescence}:

\begin{equation} \label{rs}
  r_{1,2}^{(s)}=\frac{k_z^{(1)}-k_z^{(2)}}{k_z^{(1)}+k_z^{(2)}}
\end{equation}

\begin{equation} \label{rp}
  r_{1,2}^{(p)}=\frac{\frac{\epsilon_2}{\epsilon_1}k_z^{(1)}-k_z^{(2)}}{\frac{\epsilon_2}{\epsilon_1}k_z^{(1)}+k_z^{(2)}}
\end{equation}

\noindent
and $k_z^{(1)}$, and $k_z^{(2)}$ are z-components of the the plane wave vectors in medium 1 and 2.

Then, from Equations.~\ref{rs} and~\ref{rp}, and substituting $k_z^{(1)}$, $k_z^{(2)}$, and $\frac{n_2}{n_1}$ with $k^{(1)}\cos\theta_{e1}$, $k^{(1)}\cos\theta_{e2}$, and $n$, respectively we get:

\[r_{1,2}^{(s)} =\frac{k_z^{(1)}-k_z^{(2)}}{k_z^{(1)}+k_z^{(2)}}
=\frac{k^{(1)}\cos\theta_{e1}-k^{(2)}\cos\theta_{e2}}{k^{(1)}\cos\theta_{e1}+k^{(2)}\cos\theta_{e2}}
=\frac{k^{(1)}\cos\theta_{e1}-k^{(2)}\sqrt{1-\sin^2\theta_{e2}}}{k^{(1)}\cos\theta_{e1}+k^{(2)}\sqrt{1-\sin^2\theta_{e2}}}\]

\[=\frac{k^{(1)}\cos\theta_{e1}-k^{(2)}\sqrt{1-(\frac{n_1}{n_2})^2\sin^2\theta_{e1}}}{k^{(1)}\cos\theta_{e1}+k^{(2)}\sqrt{1-(\frac{n_1}{n_2})^2\sin^2\theta_{e1}}}
=\frac{k^{(1)}\cos\theta_{e1}-k^{(2)}\sqrt{1-\frac{1}{n^2}\sin^2\theta_{e1}}}{k^{(1)}\cos\theta_{e1}+k^{(2)}\sqrt{1-\frac{1}{n^2}\sin^2\theta_{e1}}}\]

\begin{equation}
r_{1,2}^{(s)}=\frac{k^{(1)}\cos\theta_{e1}-k^{(2)}\sqrt{1-\frac{1}{n^2}+\frac{1}{n^2}\cos^2\theta_{e1}}}{k^{(1)}\cos\theta_{e1}+k^{(2)}\sqrt{1-\frac{1}{n^2}+\frac{1}{n^2}\cos^2\theta_{e1}}}
\end{equation}

\noindent
where $\theta_{e1}$ and $\theta_{e2}$ are the angles between normal vector of surface plane and $k^{(1)}$ and $k^{(2)}$ wave vectors. Finally we have $r_{1,2}^{(s)}(\nu)$ and $r_{1,2}^{(p)}(\nu)$ by replacing ${k_z^{(1)}}/{k^{(1)}}$ with $\cos\theta_{e1}$, where $\nu=\cos\theta_{e1}$ \cite{drexhage1970influence}. 

We also explored a third approach to calculate the radiative power, using a first-order Taylor series~\cite{lukosz1977fluorescence} of total radiative power in the regime that \(z_0 \ll\lambda_1\).

\begin{figure}[ht]
    \centering
    \includegraphics[width=\textwidth]{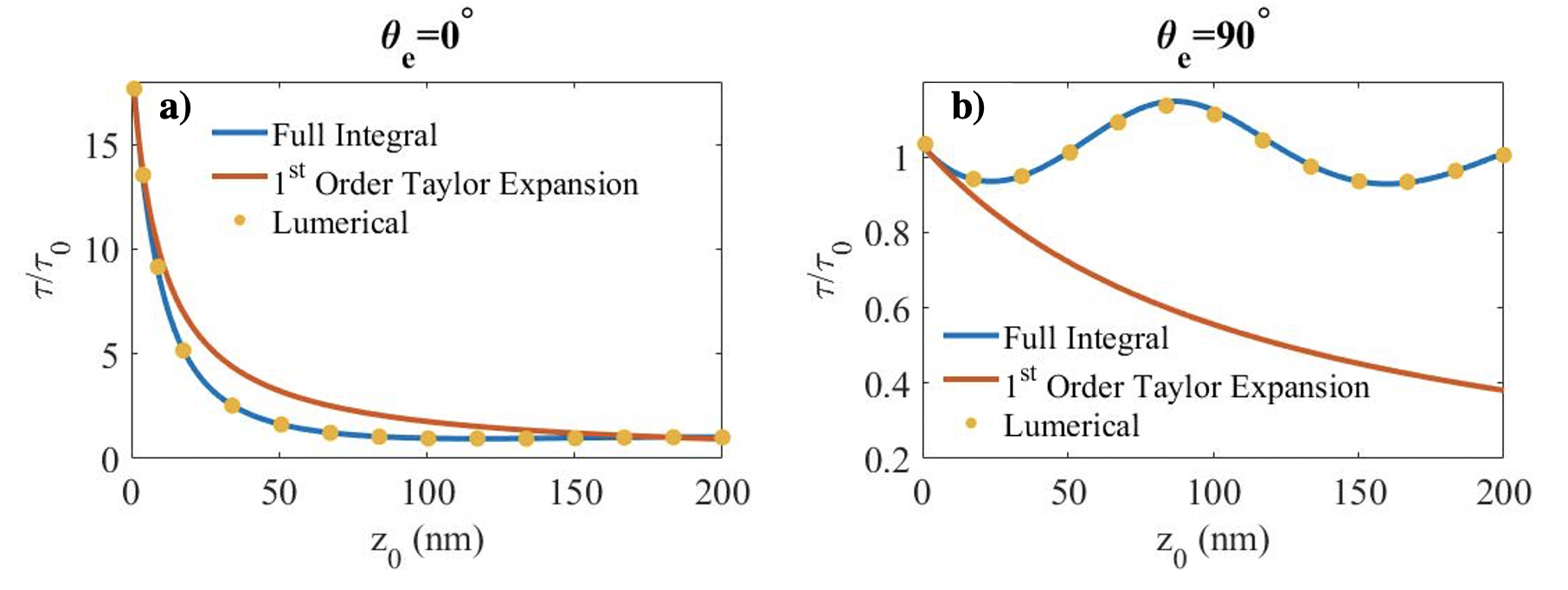}
    \caption{Normalized lifetime of an electric dipole, a) perpendicular ($\theta_e=0^\circ$), and b) parallel ($\theta_e=90^\circ$) to the interface.}
    \label{theory_lumeircal_taylor}
\end{figure}

Figure~\ref{theory_lumeircal_taylor} plots the computed lifetimes by the three approaches (FDTD, full numerical integration, and first-order Taylor expansion).
The depth-dependent lifetimes are in great agreement between FDTD and numerical integration methods. However, our results indicate that first order Taylor expansion only works well for a limited range of $z_{0}\leq 6$ nm (for \(\lambda\)=$637$ nm and $n_1=2.4$). The results in the main text are thus obtained using numerical integration. 

\subsection {Calculation of radiative powers} \label{radiative power ratio}
The spectrally averaged power ratios for the \(X\) and \(Y\) dipole transitions are calculated from Eqs. 9 and 10 as follows:

\begin{equation}
    \langle[L(z_0,\alpha,\theta_m)/L_{\infty}]\rangle_X=\sin^2(\alpha) \sin^2(\theta_m)\langle[L(z_0)/L_{\infty}]_{\perp}\rangle+[1-\sin^2(\alpha) \sin^2(\theta_m)]\langle[L(z_0)/L_{\infty}]_{\parallel}\rangle
\end{equation}

\begin{equation}
\begin{split}
    \langle[L(z_0,\alpha,\theta_m)/L_{\infty}]\rangle_Y & =\sin^2(\alpha+\frac{\pi}{2}) \sin^2(\theta_m)\langle[L(z_0)/L_{\infty}]_{\perp}\rangle+[1-\sin^2(\alpha+\frac{\pi}{2}) \sin^2(\theta_m)]\langle[L(z_0)/L_{\infty}]_{\parallel}\rangle\\
    & =\cos^2(\alpha) \sin^2(\theta_m)\langle[L(z_0)/L_{\infty}]_{\perp}\rangle+[1-\cos^2(\alpha) \sin^2(\theta_m)]\langle[L(z_0)/L_{\infty}]_{\parallel}\rangle
\end{split}
\end{equation}

\begin{equation}
\begin{split}
    \langle[L(z_0,\alpha,\theta_m)/L_{\infty}]\rangle_{\frac{X+Y}{2}}=\frac{1}{2}\sin^2\theta_m \langle [L(z_0)/L_{\infty}]_{\perp}\rangle+\frac{1}{2}(1+\cos^2\theta_m)\langle[L(z_0)/L_{\infty}]_{\parallel}\rangle
\end{split}
\end{equation}

\subsection{Shallow NV lifetimes in different media} \label{Comparison-oil-air}
\begin{figure}[h]
    \centering
    \includegraphics[width=\textwidth]{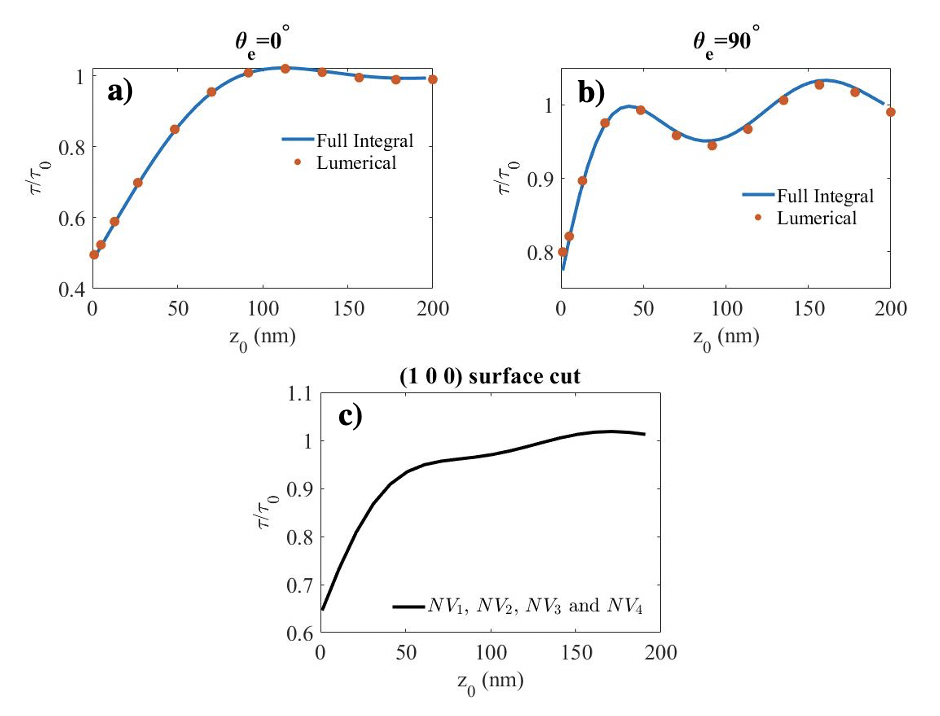}
    \caption{Normalized lifetime of an electric dipole, for $n>1$, with $n_{1}=2.4$, and $n_{2}=3.4$ at $\lambda=637$ nm a)  perpendicular, and b) parallel to the interface. c) Spectrally averaged normalized lifetime of NV in a (100) surface cut diamond versus $z_{0}$.}
    \label{theory_lumeircal_n_greater_than_1}
\end{figure}
\begin{figure}[h]
    \centering
    \includegraphics[width=\textwidth]{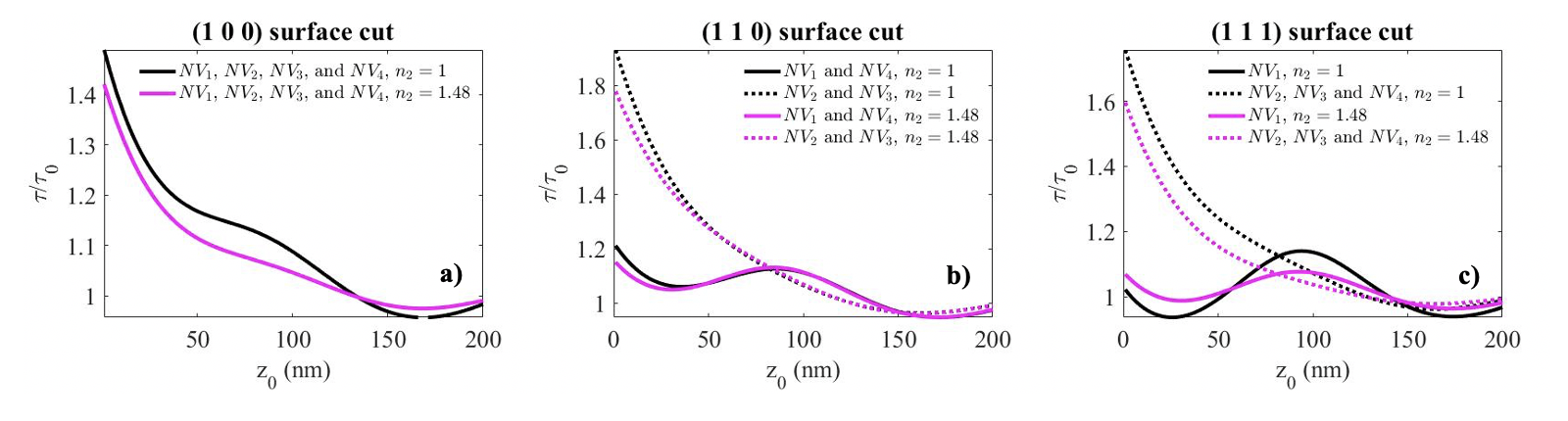}
  \caption{Comparison of NV lifetime in different $n_2$ medium in a) (100), (b) (110), and (c) (111) diamond surface cuts }
    \label{All_Tau_air_oil}
\end{figure} 

\end{document}